# A Robust and Precise ConvNet for small non-coding RNA classification (RPC-snRC)

Muhammad Nabeel Asim[a,*], Muhammad Imran Malik[b], Andreas Dengel[a], Sheraz Ahmed[a]

[a]*German Research Center for Artificial Intelligence (DFKI), 67663 Kaiserslautern, Germany*
[b]*National University of Sciences and Technology, Islamabad, Pakistan*

## Abstract

Functional or non-coding RNAs are attracting more attention as they are now potentially considered valuable resources in the development of new drugs intended to cure several human diseases. The identification of drugs targeting the regulatory circuits of functional RNAs depends on knowing its family, a task which is known as RNA sequence classification. State-of-the-art small non-coding RNA classification methodologies take secondary structural features as input. However, in such classification, feature extraction approaches only take global characteristics into account and completely oversight co-relative effect of local structures. Furthermore secondary structure based approaches incorporate high dimensional feature space which proves computationally expensive. This paper proposes a novel Robust and Precise ConvNet (RPC-snRC) methodology which classifies small non-coding RNAs sequences into their relevant families by utilizing the primary sequence of RNAs. RPC-snRC methodology learns hierarchical representation of features by utilizing positioning and occurrences information of nucleotides. To avoid exploding and vanishing gradient problems, we use an approach similar to DenseNet in which gradient can flow straight from subsequent layers to previous layers. In order to assess the effectiveness of deeper architectures for small non-coding RNA classification, we also adapted two ResNet architectures having different number of layers. Experimental results on a benchmark small non-coding RNA dataset show that our proposed methodology does not only outperform existing small non-coding RNA classification approaches with a significant performance margin of 10% but it also outshines adapted ResNet architectures.

*Corresponding author
*Email address:* Muhammad_nabeel.asim@dfki.de (Muhammad Nabeel Asim)
*URL:* malik.imran@seecs.edu.pk (Muhammad Imran Malik), andreas.dengel@dfki.de (Andreas Dengel), sheraz.ahmed@dfki.de (Sheraz Ahmed)

## 1. Introduction

Ribonucleic acid (RNA) serves as a coding template in the creation of proteins, performs various biological functions, and is largely responsible for several diseases such as Alzheimer, cardiovascular, Cancer, and type 2 diabetes [1, 2]. Primarily, RNA is classified as protein coding or non-coding. About 3% region of RNA which produces proteins is called protein coding or messenger RNAs (mRNA) while other 97% portion of RNA is known as non-coding (ncRNA) or functional RNAs [3]. Functionality of almost all protein coding RNAs is pretty much known, and studied extensively over the period because of its necessary protein encoding function, and active participation in different biological processes including embryonic development [4]. Non-coding RNAs were considered useless for quite some time, however with the advancements in biological research, it was extrapolated lately that most of the ncRNAs perform multifarious essential biological processes like dosage compensation, genomic imprinting, and cell differentiation [5, 6]. With the passage of time, analysis of ncRNA has become even more interesting because of their importance in understanding the phenomena behind human diseases and achieving stable health [5].

Non-coding RNAs differ from each other in terms of length, conformation, and biological cell function. As shown in Figure 1, ncRNAs are typically classified as small non-coding RNAs (sncRNA) or long non-coding RNAs (lncRNA). The long non-coding RNAs are greater than 200 bp in size [6] and are fur-ther splitted into linear RNAs and circular RNAs. Linear RNAs are considered noteworthy resources in gene transcription and translation [7]. Circular RNAs have the capability of gene regulation and are strongly connected with complex human diseases like lung cancer and are considered important in identification and treatment of tumours [8], [9]. Over the last decade, researchers have proposed various machine and deep learning based methodologies to classify RNAs since verifying humongous transcriptomes through manual experimentation is time consuming and an extremely expensive task [10]. Extensive research has been done to differentiate protein coding RNA from non-coding RNA. Especially, in order to discriminate long non-coding RNA (lncRNA) from protein coding or to classify long non-coding RNA (lncRNA) into their corresponding families, diverse machine and deep learning based methodologies have been proposed [11, 12, 13, 14, 7, 8].

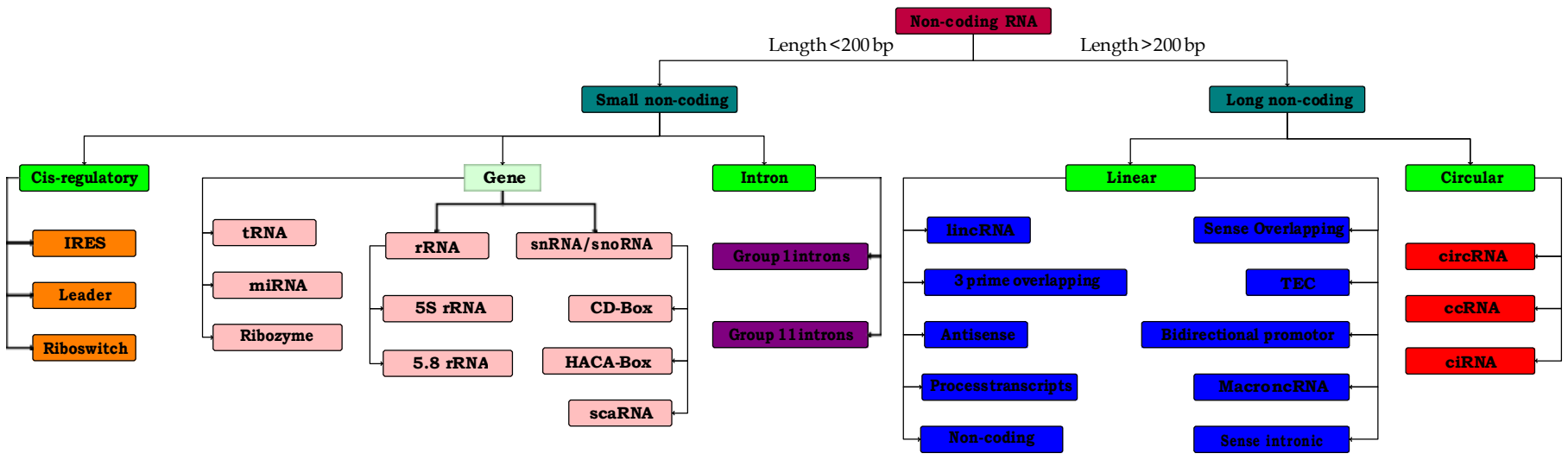

Figure 1: Overall taxonomy of non-coding RNA families, adapted from [5]. Where the magneta colored square at top layer represents non-coding RNA, dark green squares at second layer namely small non coding and long non coding refer to the subclasses of non-coding RNA. Similarly, light green squares are the major subclasses of small, and large non-coding RNA. At last level, yellow squares show the types of Cis-regulatory sequences, light pink show the kinds of Gene sequences, purple shows the subtypes of Intron sequences. On the other hand, navy blue reveal the kinds of Linear long non-codinf RNA sequences, and Circular subclasses are shown by Red squares.

The small ncRNAs possess length around 20-30 bp and are involved in translation, splicing, and regulation of genes [15]. Primarily, small ncRNAs are classified into 13 subclasses where each subclass has distinct medical and biological significance. For instance, ScaRNAs (most of which are functionally and structurally identical to snoRNAs and few are considered composite of HACA-Box and CD-Box) can guide modifications in pseudouridylation and methylation. Likewise, SnoRNA and miRNA play substantial role in cancer development [16]. In addition, miRNAs perform post transcriptional gene expression regualtion and RNA silencing. miRNAs target almost 60% of human genes and play an indispensable role in several biological processes like cell differentiation, proliferation, and death [17], [18], [19], [20]. Studies have proved that miRNAs are involved in diverse and complex human diseases such as cancer, autoimmune, cardiovascular, and neurodegenerative diseases [21]. Similarly, Ribosomal RNA (rRNA) is essential for all living organisms. It has an essential role in protein synthesis and its characteristics are considered extremely valuable for the development of multifarious antibiotics. In addition, 5S ribosomal-another kind of rRNA, exists in ribosome. Although its function is not discovered yet, it has been seen that their deletion substantially alleviate protein synthesis and also produce detrimental effects on fitness of the cell [22]. Likewise, 5.8S ribosomal RNA actively participates in protein translocation [23]. It forms covalent connection with tumour suppressor proteins [24], can be used to detect miRNA [25] and understand other rRNA pathways and processes in the cell [26].

Classification of small non-coding RNAs (sncRNAs) becomes substantial because of their large number and distinct functions. It can help biologists and clinicians to better understand the impact of sncRNAs in the development of various diseases and biological operations. Furthermore, classification of sncRNAs is also important in developing cancer therapeutic strategies [5]. According to our best knowledge, two computer based (in particular, deep learning based) approaches have reported the highest classification results for small non-coding RNA to date. The first approach- " nRC" by Antonino Fiannaca et al. [27]- comprises of three fundamental tasks including estimation of secondary structures from Rfam dataset (publicly available benchmark dataset containing 8920 samples belonging to 13 sncRNA subclasses), extraction of common substructures, and classification into 13 known ncRNA classes using a convolutional neural network (CNN). This approach achieved 81% ncRNA classification accuracy. The second approach- proposed by Emanuele Rossi et al. [28]- extracts sec-ondary structural features from the same Rfam database. However, rather than using simple convolutional neural network, they utilize Graph based convolutional architecture for the extraction of discriminative features and classification. According to the best of our knowledge, this approach reported the state-of-the-art performance with 85% accurate classification of small non-coding RNA sequences.

Note that the state-of-the-art small non-coding RNA classification approaches take secondary structure of RNA sequences as input and extract discriminative features by utilizing convolution layers or graph based methodologies. However, feature extraction methods based on secondary structures usually only consider the global characteristics while ignoring the mutual influence of the local struc- tures [29]. Such methods usually have a possibility of neglecting important information that might have been available in the primary sequences and has potentially lost while developing the secondary structures(on whom the final classification is based on). Furthermore, secondary structure based methods in- tegrate high-dimensional feature space which is computationally inefficient [29]. In the current paper, rather than extracting discriminative features from any secondary structures, we propose to use the primary RNA sequences directly. We present a Robust and Precise Convolutional neural network for small non- coding RNA Classification (RPC-snRC) system. The proposed system is based on an end to end small non-coding RNA classification methodology which uses a set of deep convolutional layers for the extraction of discriminative features by utilizing positioning and occurrences information of various nucleotides in RNA sequences. To evaluate the integrity of proposed methodology, we perform ex- periments on the publicly available benchmark dataset provided by Antonino Fiannaca [27]. The proposed system clearly outperforms all the existing meth- ods and outshines the previous state-of-the-art method (by Emanuele Rossi et al. [28]) by a fair 10% margin in terms of different performance metrics including accuracy, precision, recall and $F$ 1-measure. In addition, extensive experimen- tation is performed with different sequence k-mers (1-mer, 3-mers) and repre- sentation schemes including one hot vector, random embedding initialization, and pre-trained prot2vec embeddings so that it could be concluded weather deep architectures performs better at atom level or word level and which kind of feature representation is better for desciminative feature extraction. Moreover, to further analyse the idea of utlizing primary RNA sequences, we perform experiments with two adapted deep ResNet architectures which vary in terms of hyper-parametrs. Both of these architectures also outperform the previous state-of-the-art deep learning approaches thereby validating the idea of utilizing the primary RNA sequences for classification.

## 2. Related Work

Non coding Ribonucleic Acid (ncRNA) has been classified into a range of distinct classes or families which vary in function and composition. The interest to develop sophisticated methods for ncRNA classification has rocketed over the period since knowing the family of ncRNA is substantial for drug targeting and understanding growth of various complex diseases. Non-coding RNA classification is a vast domain where classification at different levels of ncRNA(shown in figure 1) has been performed. Mainly, researchers have been focusing to 1) distinguish non-coding RNA from coding RNA, 2) categorize ncRNA into long and small non-coding RNA , 3) segregate non-coding RNA into its subtypes such as circular RNA, and to 4) classify small non-coding RNA into its 13 sub-classes. Classification at each level facilitates distinct biological advantages. In order to discover more classes of non-coding RNA, researchers have also developed clustering-based computational methodologies. Although the main focus of this paper is small non-coding RNA classification, however considering the importance of ncRNA classification, this section provides an overview of state-of-the-art non-coding RNA classification approaches at diverse levels. It also sheds light on clustering-based approaches for non-coding RNA identification

In the last decade, researchers were more inclined towards the development of computational methodologies which can discriminate between non-coding RNA and coding RNA. Peter et al. [30] proposed a method, namely RNAz, based on Support Vector Machines (SVM) to classify ncRNAs. The RNAz combined sequence analysis approach with structure prediction. Primarily two components consensus secondary structure and thermodynamic stability were used. RNAz also integrated multifold sequence alignment and pairwise alignment of ncRNA sequence with extremely high sensitivity and specificity. They utilized RFAM genomic database [31] containing ncRNAs of humans, mice, zebrafish, and rats. RNAz exploited RNA folding of least free energy and computed z-scores by performing regression through SVM. Input parameters of proposed approach were number of alignment sequences, structure conservation index (SCI), and the mean of MFE z-score [32] of diverse sequences present in alignment excluding gaps. It also utilized the functionality of program namely RNAALIFOLD [33] which was primarily developed to estimate secondary structure from aligned sequence. RNAz used a folding algorithm to predict the secondary structure of RNA' s through implementing dynamic, and robust programming algorithms. They reported that when SCI was almost zero, it indicated that consensus structure was not found by the RNAALIFOLD, contrarily perfect conserved structures had the SCI of almost 1. RNAz produced decent results for genomic annotation performed at large scale. Likewise, Jinfeng at al. [34] presented a method based on SVM namely Coding or non-coding (CONC) to classify ncRNAs. It integrated multiple sequence alignment and used the databases FANTOM3[35], NONCODE[36], and RNAdb[37] for experimentation. This method utilized composition of amino acid, exposed residues estimated percentage, peptide length, compositional entropy, found homologs from mentioned databases searches, alignment entropy, and estimated content of secondary structure.

In order to raise the performance of ncRNA classification further, few researchers explored ensemble approaches considering the effectiveness of decision trees. For instance, Marasri et al. [38] came up with a hybrid tool for the task of ncRNAs classification. They combined an ensemble of several decision trees and random forest with logstic regression model to discriminate short, and long ncRNA sequences. This tool includes naive feature SCORE which was computed by logistic regression through the combination of five features, i.e., structure, robustness, sequence, modularity, and coding potential. For experimentation, it used multiple datasets including, RefSeq [39], Rfam [31], lncRNAdb [40], and genome database "GenBank" of NCBI. In the proposed methodology, a set of 369 features were extracted to predict ncRNAs. Amongst these features, discriminative features were acquired through feature selection based on correlation and genetic algorithm. While logistic regression was utilized to locate relationships among features, sequence similarity was facilitated by fundamental local alignment finder (BLAST) [41]. Random forest acted as primary classifier. Ensemble of several decision trees in random forest was capable to acquire heterogeneity of ncRNA subfamilies. This methodology was robust as it exploited composite features which raised the classifier performance. This approach was used to classify known ncRNAs, and also unknown ncRNAs. Similarly, Yanni et al. [42] presented a method namely lncRNA-ID based on balanced decision trees to identify long ncRNAs. This method utilized multiple sequence alignment and LncRNADisease database[43] for experimentation.

Furthermore, researchers also experimented with unsupervised methodologies for ncRNAs identification. For example, Yasubumi et al. [44] presented a methdology, namely EnsembleClust, for hierarchical clustering of ncRNAs. This methodology enabled the discovery of new ncRNA families [44] and aided to investigate functional diversity of ncRNAs. EnsembleClust implemented an unsupervised approach which utilized unlabelled data to construct clusters of ncRNAs on the basis of structural alignment results. As the computation of structural alignment was extremely expensive, approximate algorithms were utilized which considered all possible secondary structures and sequence alignments. In addition, for the sake of accurate clustering, a robust measure was used which considered primary sequences, and secondary structures. EnsembleClust produced better performance when compared with previous approaches such as FOLDALIGN [45], Stem kernel [46], and LocARNA [47]. Moreover, Milad et al. [48] came up with an approach, RNAscClust, to identify ncRNAs. RNAscClust was used to combine RNA sequences through structure conservation, and graph oriented motifs [49]. This approach used structural similarities in order to group paralogous RNAs. RNAscClust enbaled clustering of humongous occurrences. Sequences were transformed into a graph, where every nucleotide was taken as graph vertices represented with the labels A, U, G, C in form of base pair connections, and the edges were representing encoded backbone. The structures were compared with one another through graph kernels. This method considered the changes of base pairs-which were never encountered by previous clustering approaches. For experimentation, Rfam database having ncRNA sequences was used. Authors reported that the proposed method managed to facilitate accurate clustering which made it possible to align large clusters efficiently.

Considering the promising performance of deep neural network for diverse natural language processing tasks, researchers employed Convolutional Neural Networks (CNNs) to classify ncRNAs. For example, Yasubumi et al. [50] proposed a methodology CNNClust to make the clusters of ncRNAs. This technique integrated pair wise alignment of ncRNA sequences. CNN was trained using positional weigh matrices of underlying sequence motifs. Two kinds of neural word embeddings, one hot encoding and word2vec, were used by CNNClust. Information of secondary structures and read mapping were also utilized in CNNClust. Matrix of similarity score was computed for each pair of RNA sequences and clustering was performed to group highly similar structures. CNNClust categorizes ncRNA into either positive or negative class. When both ncRNA sequences belong to the same class then it was classified as positive otherwise negative. Several new kinds of ncRNA such as microRNA, tRNA, and snoRNA were discovered through this approach. For experimentation, authors used Rfam, HUGO gene nomenclature committee (HGNC), and Genomic tRNA (GtRNAdb) datbabses. Similarly, Antonino et al. [51] presented an approach, nRC, for classification of ncRNAs. This approach used features of secondary structures and incorporated alignment of multiple sequences to categorize 13 known ncRNA classes using a CNN. The nRC utilizes IPKnot43 which is capable of predicting secondary structures and generate an accurate graph based on multifarious topologies of non-coding RNA sequences. IPKnot43 yields a graph database as it generates an undirected label graph for every input transcript. Considering the ideology that graphs having similar substructure usually belong to same RNA family, common subgraph extraction is exploited with a minimum threshold to locate frequent substructures which represent features of diverse small non-coding RNA subclasses. In order to locate subgraphs, nRC utilizes Molecular Substructure Miner (MoSS) which produces common subgraphs using a depth-first search. In this way, nRC only considers close common subgraphs. Lastly, nRC leverages the power of a CNN containing two convolutional and two fully connected layers.

Likewise, few researchers experimented with Recurrent Neural Networks (RNN) to classify ncRNAs. Sungroh et al. [52] presented a methodology namely lncRNAnet for the ncRNA classification. LncRNAnet identified long ncRNA through next generation sequencing [52] and deep learning. They used both CNN and RNN. While RNN was exploited to model RNA sequences, CNN was used to spot stop condons in order to locate an idicator of open reading frame. LncRNAnet showed decent performance while classifying short length RNA sequences. It learned intrinsic features through RNN for modelling RNA sequences. Authors performed experimentation on GENCODE, ENSEMBL and Human and Vertebrate Analysis and Annotation (HAVANA) databases. They reported that the proposed methodology produced robust performance regardless of variable sequence length, and helped to identify latest lncRNA from large transcriptome data. Moreover, Sungroh et al. [53] proposed a methodology based on deep RNN for the task of ncRNA classification. The proposed method utilized the features of secondary structures to identify ncRNA and incorporated pairwise alingment of sequences. Authors used fRNAdb, NONCODE, and NCBI datasets for extensive experimentation.

Contrarily deep sequencing has also been employed for ncRNA classification. For instance, Yasubumi et al. [54] presented an approach SHARAKU based on deep sequencing for ncRNA classification. SHARAKU incorporated an algorithm which aligned read mapping profiles of ncRNAs next generation data containing sequences. This system also implemented a program for the alignment of read mapping profile which used decomposition for the sake of folding and aligning RNA sequences at the same time [55]. Profiles of read mapping allowed the detection of common patterns. Secondary structure and sequence information were acquired concurrently in this approach. The proposed approach helped to locate ncRNAs specifically combined in brain. The authors used NCBI, ENSEMBLE, and next generation sequencing output databases as reference. SHARAKU managed to achieve better performance than deepBlock-Align [56]. Likewise, Rosemarie et al. [57] presented a method based on next generation deep sequencing for the task of classifying ncRNA. The proposed method utilized features of protein coding to discriminate among coding and non-coding RNAs. It incorporated alignment of pairwise sequences and used lncRNA, NONCODE, NCBI datasbases for experimentation.

In order to improve the performance of small non coding RNA classifica- tion, more recently, Emanuele Rossi [28] proposed Graph convolutional neural network based methodology which also takes secondary structural features as input. This methodology uses Graph convolutions for the extraction of discrim- inative features from the secondary structural features. According to our best knowledge, this is the latest methodology which has excluded manual feature en- gineering and produced state-of-the-art performance for small non-coding RNA classification. Table 1 summarizes the state-of-the-art work for RNA sequence classification.

In this paper, we proposed a RPC-snRC methodology which takes input RNA sequence data and utilises convolutional layers for the extraction of discriminative features which are eventually passed to dense layers for classification. Note that our methodology does not require any alignment or manual feature extraction technique as it provides an end to end deep learning system which takes RNA sequences as input and provides class label as output.

| Method | Database | Alignment | Features | Technique | Categorization type |
|---|---|---|---|---|---|
| RNAz [58] | Rfam | Pairwise and multiple sequence alignment | Thermodynamic stability measure, consensus secondary structure | SVM | Classification into coding or non coding RNA |
| CONC [59] | RNAdb, NONCODE, FANTOM | multiple sequence alignment | Amino acid composition, peptide length, predicted secondary structure content, predicted percentage of exposed residues, compositional entropy, number of homologs from database searches and alignment entropy | SVM | Classification into coding or non coding RNA |
| Hybrid random forest [60] | Rfam, RefSeq, NCBI GenBank genome database and lncRNAdb database | multiple sequence alignment | sequence, structure, structural robustness, modularity and coding potential | Random forest | Classification into small non coding or long non coding RNA |
| Deep RNN [61] | NCBI, fRNAdb, NON-CODE | pairwise sequence alignment | Secondary sequence features | RNN | Micro RNAs Identification |
| lncRNAID [62] | LncRNADisease database | profile hidden Markov model (profile HMM)-based alignment | open reading frame (ORF), protein conservation and ribosome interaction | Random forest | Long non coding RNA identification |
| lncRNAnet [62] | GENCODE, ENSEMBL and Human and Vertebrate Analysis and Annotation group databases | multiple sequence alignment | Open reading frame (ORF) indicator | RNN | Long non coding RNA identification |
| EnsembleClust [63] | ENSEMBLE | Pairwise sequence alignment | structural alignments score | Hierarchical Clustering | Clustering of non coding RNA |
| RNAscCLust [64] | Rfam | - | structure conservation and graph-based motifs | Hierarchical Clustering | Clustering of non coding RNA |
| SHARAKU [65] | NCBI Reference sequence database, ENSEMBLE database and next generation sequencing output | Pairwise sequence alignment | Similarity score matrix | Random forest | Clustering of non coding RNA |
| CNNClust [66] | Rfam, HUGO gene nomenclature committee (HGNC) databases, Ensembl and genomic tRNA database | Pairwise sequence alignment | Derived position weight matrices of sequence motifs | CNN | Clustering of non coding RNA |
| Deep next generation sequencing [67] | NONCODE, NCBI, lncRNA | Pairwise sequence alignment | Protein coding features | Deep next generation sequencing | Classification into coding or non coding RNA |
| circ-Deep [68] | CircRNADb | - | RCM features, conservation features | CNN and LSTM | Long non coding Circular RNA classification |
| nRC [69] | Rfam | Multiple sequence alignment | Secondary structure features | CNN | Classification of small non coding RNA |
| RNAGCN [28] | Rfam | Multiple sequence alignment | Secondary structure features | graph convolutional network | Classification of small non coding RNA |

Table 1: Summary of the previous work for non-coding RNA classiftcation and clustering in terms of exploited technique, alignment of sequences information, and type of features used as an input. In Table three methodologies namely Ensemble clust, RNAscClust, SHARAKU and CNN clust makes groups of non coding RNAs according to their structural similarities. LNCRNAid and LncRNANet performs long non coding RNA identiftcation. Two machine learning based methodologies namely RNAZ, and CONC differentiates between coding and non coding RNA sequences. Hybrid random forest methodology is used to discriminate between small non coding and long non coding RNA sequences. Deep RNN methodology identiftes microRNAs.Two deep learning methodologies namely nRC and RNAGCN methodologies performs classiftcation of small non coding RNA.

## 3. Materials and methods

Following the triumph of deep learning methodologies in the ImageNet Large Scale Visual Recognition Challenge (ILSVRC), [1], researchers were more interested to employ deep learning for diverse computer vision, natural language processing, and bioinformatics tasks [70], [71], [72], [73], [74]. Generally, the aim was to develop deeper architectures with proper gradient flow among the layers which could learn better hierarchical representation of features.

[1] http://www.image-net.org/challenges/LSVRC/

Deoxyribonucleic acid (DNA) and ribonucleic acid (RNA) sequences are often treated in the same way as traditional text is treated in natural language processing [75]. A term K-mers is used for DNA and RNA sequences where a group of three or four nucleotides are combined to form a word known as 3-mers or 4-mers. Today, however, there is a debate about which atom-level (single nucleotide known as character or K-mers known as word) would be the most effective representation for DNA and RNA sequence analysis tasks? Furthermore, researchers are also working on protiomic and genomic data to provide biomedical pretrained neural word embeddings for different k-mers. This is because neural word embeddings, known as continuous representations of features or words, have played an important role to improve the performance of various NLP tasks. In this regard, recently Asgari et al. [76] provided pretrained neural word embedings for proteins and genes. However, there are again several open questions about the impact of utilizing pretrained neural word embedings for DNA, and RNA sequence analysis, e,g., will deep architectures learn better features using pretrained word embedings of protines and genes?

This paper presents a robust and precise convnet based system for small non-coding RNA classification.The proposed system takes direct RNA sequence data as input and utilises convolutional layers for the extraction of discriminative features which are eventually passed to dense layers for classification. This system does not require any alignment or manual feature extraction provides an end to end deep learning based system which takes primary RNA sequences as input and provides class label as output. Furthermore, to provide answers of above questions, we have performed detailed experimentation on small non-coding RNA classification dataset with the proposed system and also with two adapted Res-Net architectures. Parameter details of the adapted Res-Net architectures are summarized in Table 2.

| Layer_Name | Res18_nRC | | Res50_nRC | |
|---|---|---|---|---|
| | Output Size | Parameters detail | Parameters detail | Output Size |
| Conv-1 | 64×1182 | (64,3), s=1,p=1 | | 64 x 1182 |
| Conv-2 | 64×1182 | $\begin{bmatrix}(64,17)\\(64,17)\end{bmatrix}\times 2, p=8$ | $\begin{bmatrix}(64,1), p=0\\(64,17), p=8\\(256,1), p=0\end{bmatrix}\times 3$ | 256 x 1182 |
| Pool-1 | 64×591 | (2, 2) | | 256 x 591 |
| Conv-3 | 128×296 | $\begin{bmatrix}(128,17)\\(128,17)\end{bmatrix}\times 2, s=2, p=8$ | $\begin{bmatrix}(128,1), p=0\\(128,17), p=8\\(512,1), p=0\end{bmatrix}\times 4, s=2$ | 512 x 296 |
| Pool-2 | 128×148 | (2, 2) | | 512 x 148 |
| Conv-4 | 256×74 | $\begin{bmatrix}(256,17)\\(256,17)\end{bmatrix}\times 2, s=2, p=8$ | $\begin{bmatrix}(256,1), p=0\\(256,17), p=8\\(1024,1), p=0\end{bmatrix}\times 6, s=2$ | 1024 x 74 |
| Pool-3 | 256×37 | (2, 2) | | 1024 x 37 |
| Conv-5 | 512×19 | $\begin{bmatrix}(512,17)\\(512,17)\end{bmatrix}\times 2, s=2, p=8$ | $\begin{bmatrix}(512,1), p=0\\(512,17), p=8\\(2048,1), p=0\end{bmatrix}\times 3, s=2$ | 2048 x 19 |
| Pool-4 | 512 x 9 | (2, 2) | | 2048 x 9 |
| Output | 13 | Flatten-4608 | Flatten-18432 | 13 |

Table 2: Architecture summary of Res18-nRC and Res50-nRC: In both architectures, before res modules, there is a convolutional layer through which ncRNA samples are passed. Both architectures have 4 res modules, while each module of Res18-nRC has 2 basic blocks, where each basic block has two convolutional layers, but Res50-nRC architecture has variable bottleneck blocks in each res module which are mentioned by a number out side the matrix brackets, i.e., ftrst res module has 3 bottleneck blocks and second has 4. In ftst matrix (64,17) 64 represents number of feature maps and 17 shows the kernel size.

### *3.1. Proposed Methodology*

This section briefly describes the proposed methodology of RPC-snRC for classification of small non-coding RNA. We develop a deep classifier in which a phenomena similar to DenseNet is used to enable proper flow of gradient between the layers. RPC-snRC utilizes a set of convolutional layers for extraction of discriminative features from the primary sequences of small non-coding RNA. Discriminative features are then fed to dense layers for classification of sequences into set of predefined classes.

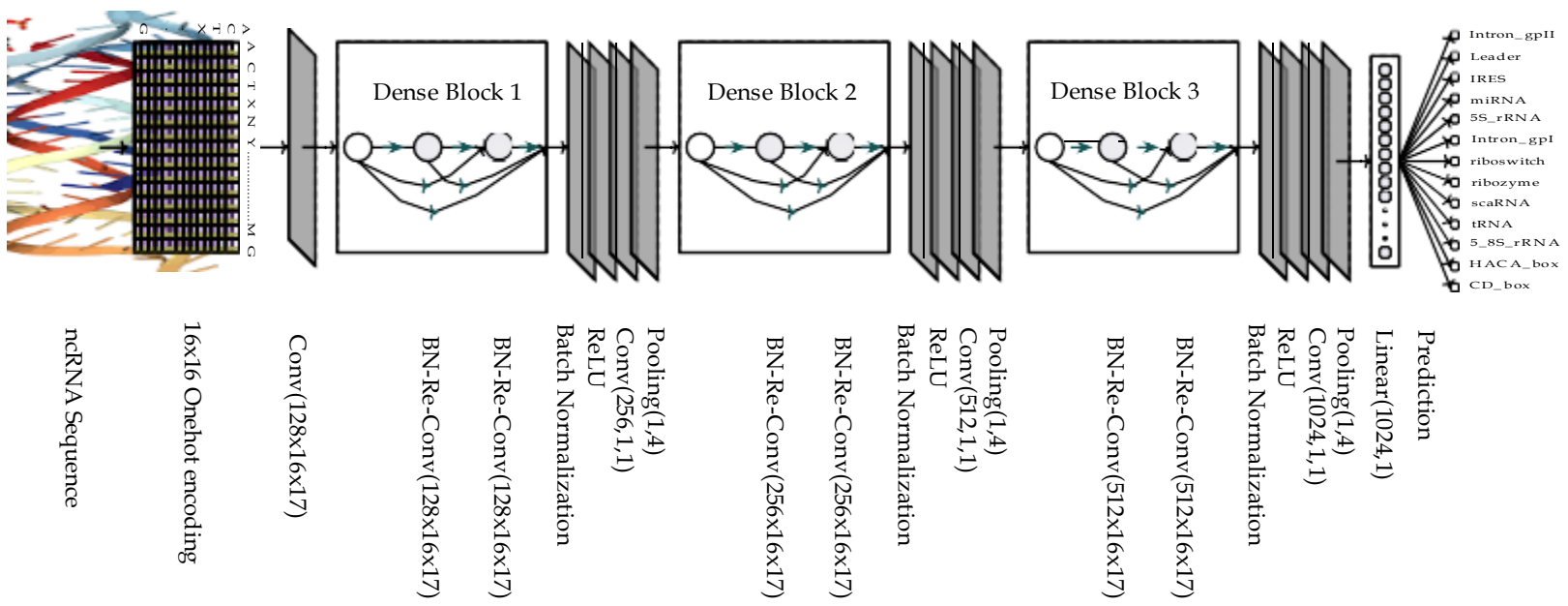

Figure 2: Proposed RPC-snRC methodology for small non-coding RNA classiftcation. In ftgure, (128,16,18) indicates there are 128 kernels, each of width 16 and length 18 in a convolutional layer and (1,4) indicates kernel width and length are set to 1 and 4 respectively in a pooling layer. Others have the similar meaning.

Figure 2 illustrates the architecture of proposed methodology along with noteworthy model parameters. The proposed RPC-snRC methodology is based on three dense modules. Each dense module contains the same number of layers, however output units get doubled in every following dense module. Each dense module first performs batch normalization on the given input and then it applies ReLu activation to introduce non-linearity followed by convolution operation to extract discriminative features. Finally it repeats the discussed operations one more time in order to better learn hierarchical representation of data. Each dense module is followed by a transition layer which performs batch normalization, ReLu activation, convolution with the filter size $1 \times 1$, and max pooling with the size of 4 to retain discriminative features and discard useless features. Dense architecture was proposed by Gao Huang et al. [74], and has been widely utilised for various applications of computer vision, we utilize this architecture for sequence data which is one dimensional and entirely different from visual data. Integral components of the proposed methodology such as DenseNets, Dense connectivity, Composite function, Pooling layers, Growth rate and Bottleneck layers which are adapted to cope one dimensional data, are discussed below.

### *3.1.1. DenseNets*

Consider a small non-coding RNA sample $S_0$ that is passed through a convolutional network. The network consists of L layers, each of which performs a non-linear conversion $H_L(\cdot)$, where $L$ indicates the layer. $H_L(\cdot)$ may be a composite function for operations like batch normalization [77], rectified linear units (ReLU)[78], Pooling [79], or Convolution (Conv). We refer to the $L^{th}$ layer output as $x_L$.

***Dense connectivity***. State-of-the-art feed-forward convolutional networks attach the $L^{th}$ layer output as an input to the $(L+1)^{th}$ layer, which produces the following transition layer $x_L = H_L(x_L - 1)$ [70]. Resnets [73] along with skip connection strategy use an identity function to bypass non-linear transformations shown in equation 1

$$X_L = H_L(X_{L-1}) + x_{L-1} \quad (1)$$

ResNets benefit is that the gradient can flow straight from subsequent layers to previous layers through the identity function. However, the identity function and output of $H_L$ are mixed by summation which can hinder the flow of data in the network.

We utilise Densenet a distinct connectivity model to further enhance the information flow between layers. In this model $L^{th}$ layer gets all previous layers ' feature maps, $x_0, \cdots; x_{L-1}$, as input.

$$X_L = H_L([x_0, x_1, \cdots; x_{L-1}]) \quad (2)$$

In equation 2, $x_0, \cdots ; x_{L-1}$ relates to the concatenation of the feature maps in the $0, \cdots, L-1$ layers

***Composite function***. Following He et al. [73], we define $H_L(\cdot)$ as a composite function of three successive operations: Batch Normalization (BN) [77], accompanied by Activation function named as rectified linear unit (ReLU)[78] and a convolution (Conv) layer.

***Transition layers***. We refer to the layers between blocks that perform convolution and pooling operations as transition layers. The procedure of concatenation used in equation 2 is not applicable if size of feature maps is variable. In our architecture we split the network into various tightly linked dense blocks to make the same size of feature maps. Down sampling is performed through transition layers which consist of batch normalization layer and a convolution layer of kernel size 1, followed by an average pooling layer of kernel size 4.

***Growth rate***. If each composite function $H_L(\cdot)$ produces $N$ feature maps, then $L^{th}$ layer will have $N_0 + N \times (L-1)$ input feature-maps, where $N_0$ denotes number of channels in the input layer. We refer to the $N$ hyper parameter as the network's growth rate.

### *3.2. Validation method and evaluation criteria*

We perform experimentation on a small non-coding RNA classification dataset manually tagged by Antonino et al. [27]. This is the only benchmark dataset which is publicly available. It consists of 8920 samples that belong to 13 different ncRNA classes: miRNA, ribozymes, 5S rRNA, 5_8S_rRNA, HACA-box, CD-box, tRNA, scaRNA, IRES, Intron_gpI, Intron_gpII, riboswitch, and leader. This dataset is quite balanced as almost every class has 700 samples except the ires class which contains 520 samples. Detailed statistics of this dataset are shown in table 3.

The dataset has benchmark defined split with 6320 training and 2600 test samples belonging to 13 classes of ncRNA. In the test set, each class has 200

| Classes | No.of Samples | Max-seq length | Min-seq length |
|---|---|---|---|
| IRES | 520 | 630 | 53 |
| Intron_gpI | 700 | 1182 | 133 |
| leader | 700 | 237 | 38 |
| scaRNA | 700 | 445 | 78 |
| S5_rRNA | 700 | 199 | 61 |
| miRNA | 700 | 631 | 52 |
| tRNA | 700 | 177 | 47 |
| riboswitch | 700 | 399 | 44 |
| ribozyme | 700 | 1136 | 41 |
| S8_rRNA | 700 | 290 | 50 |
| CD-box | 700 | 404 | 54 |
| HACA-box | 700 | 508 | 59 |
| Intron_gpII | 700 | 241 | 48 |

Table 3: Characteristics of Non-coding RNA classiftcation dataset, where Max-seq length and Min-seq length illustrate maximum and minimum length of neucleotides in each class

samples, whereas in training set, each class has 500 samples except the IRES class which has 320 samples available for training. A well known statistical cross validation method namely leave one out cross validation is used to better analyze behaviour of the proposed model. We have used the training set for training and validation of the proposed model while test set is only used for the final evaluation of the model. Furthermore, training set is split into 5 equal parts, 4 parts are used to train the model and the $5^{th}$ part is used to validate the trained model. For dual evaluation, trained model is also evaluated on test data set which was held out separate. The process of training and dual evaluation is repeated five times where every time test set remains the same but every next fold is taken as validation set. Final results are computed by taking the average of 5 results which are produced by the proposed model at each fold.

***Evaluation Metrics***. The proposed system is evaluated using four different evaluation metrics namely Accuracy, Precision, Recall, and $F_1$ measure. All four evaluation metrics compute scores by utilising four parameters, i.e., true positives, true negatives, false positives, and false negatives, as shown in Table 4.

| | Predicted Class | | |
|---|---|---|---|
| Actual Class | | Class=yes | Class=no |
| | Class=yes | True Positive | False Negative |
| | Class=no | False Positive | True Negative |

Table 4: Confusion Matrix where True Positive illustrates the count of correctly predicted positive class values, e.g., if both the actual and predicted class labels will be yes then it will be considered as true prediction of positive class label. Similarly, True Negative is accurate prediction of negative class labels. False Positive denotes the count for wrongly predicted class labels , i.e., when actual class is 'no' but model predicts 'yes' , similarly, False Negative is wrong prediction of 'no' class when actual class was 'yes' .

***Accuracy***. Accuracy is considered as a reasonable metric when dataset is symmetric-where values of false negatives and false positive are nearly equal. It computes the ratio of correctly predicted samples to the total samples.

$$Accuracy = \frac{T_p + T_n}{T_p + T_n + F_p + F_n} \quad (3)$$

***Precision***. Precision is the ratio between correctly predicted positive samples and total predicted positive samples.

$$Precision = \frac{T_p}{F_p + T_p} \quad (4)$$

***Recall***. Recall is the ratio among correctly predicted positive samples and actual number of positive samples. It is also known as sensitivity. Recall is preferred when we are more concerned with false negatives. For instance, if a person having cancer is predicted as normal then the value of false negative gets high which eventually decreases the recall.

$$Recall = \frac{T_p}{F_n + T_p} \quad (5)$$

$F_1$ ***Measure***. $F_1$ measure is harmonic average of precision and recall. It performs better than accuracy for imbalance class distributed dataset because it keeps track of both precision and recall.

$$F_1 Measure = 2 \times \frac{Precision \times Recall}{Precision + Recall} \tag{6}$$

## 4. Experimental setup and Results

We implement the proposed RPC-snRC and ResNet based methodologies in Python using Pytorch [80]. Detailed parametric description about adapted ResNet based methodologies is summarized in Table2. Cross entropy is used as a loss function with Adam [81] optimizer where learning rate is initialized from 0.001. In order to alleviate training time, an early stopping approach is used. High-performance NVIDIA GeForce GTX 1080Ti GPU is used for experimentation.

***Results***. This section briefly describes performance of the proposed RPC-snRC classification system and two adapted ResNet architectures (res-net 18 layer, res-net 50 layer) for the task of ncRNA classification. It shows the impact of three sequence representation schemes while treating RNA sequence as a set of characters, and k-mers based word for both proposed and adapted methodologies. For experimentation, we have fixed the sequence length to 1180 characters which is essential for convolution operation. To make all sequences of equal lengths, we apply padding where the size of sequence is less than 1180 and truncate extra characters in the opposite case. Experimentation is performed in two different ways: First, RNA sequence is taken as a set of characters with two different representation schemes namely one hot vector encoding and random embedding initialization, which are separately fed to the proposed RPC-snRC system.Second, we generate 3-mers of the sequence by rotating a window of size three on the sequence. K-mers based sequence representation along with one hot vector encoding, random embedding initialization, and pretrained word embeddings provided by Asgari et al. [76] are fed to the proposed RPC-snRC system.

| | RPC-snRC | | | | Res18-nRC | | | | Res50-nRC | | | State-of-the-art | |
|---|---|---|---|---|---|---|---|---|---|---|---|---|---|
| Performance Measures | Character one-hot | 3-mers one-hot | 3-mers random embeddings | 3-mers prot2vec embeddings | Character one-hot | 3-mers one-hot | 3-mers random embeddings | 3-mers prot2vec embeddings | Character one-hot | 3-mers random embeddings | 3-mers prot2vec embeddings | nRC [69] | RNAGCN [28] |
| Accuracy | **0.9538** | 0.9285 | 0.9327 | 0.9326 | 0.9169 | 0.8842 | 0.8880 | 0.9000 | 0.8680 | 0.8365 | 0.8915 | 0.7838 | 0.8573 |
| Precision | **0.9539** | 0.9312 | 0.9344 | 0.9322 | 0.9185 | 0.8859 | 0.8929 | 0.9000 | 0.8701 | 0.8377 | 0.8941 | 0.7780 | - |
| Recall | **0.9538** | 0.9285 | 0.9326 | 0.9326 | 0.9169 | 0.8842 | 0.8880 | 0.9000 | 0.8680 | 0.8365 | 0.8915 | 0.7830 | - |
| F1-Score | **0.9536** | 0.9286 | 0.9328 | 0.9319 | 0.9174 | 0.8842 | 0.8880 | 0.8987 | 0.8680 | 0.8357 | 0.8921 | 0.7790 | 0.8561 |

Table 5: Performance of the proposed RPC-snRC, Adapted (Res18-nRC, Res50-nRC), and state-of-the-art (nRC [69], and RNAGCN [28]) methodologies on the benchmark small non-coding RNA dataset.

Table 5 compares the performance of state-of-the-art and adapted res-net based methodologies with the proposed RPC-snRC methodology for the task of small non-coding RNA classification. It also illustrates the performance of the proposed RPC-snRC methodology when RNA sequence is treated as set of characters, 3-mers based features with random, and pre-trained neural word embeddings. As is depicted by the Table 5 renowned methodology proposed by Antonio Fiannaca et al. [27] managed to achieve the performance figures of 78%, 77%, 78%, and 77% in terms of accuracy, precision, recall, and F 1 measure, respectively. This performance is outperformed by a recent Graph Convolutional Neural architecture based methodology given by Emanuele RossiGet et al. [28] as it marked state-of-the-art performance for small non-coding RNA classification with 85.7% accuracy. However, the adapted ResNet-18 and Res-Net-50 just manage to produce the peak performance of 91%, and 89% by representing RNA sequences as character with one hot encoding and as 3-mers features with pre-trained prot2vec embedding, respectively. On the other hand, the proposed RPC-snRC classification system has significantly outperformed the state-of-the-art methodology as ewll as the two dapted Resnet architectures in all settings. While, RPC-snRC with 3-mers random embedding initialization and pre-trained neural word embeddings schemes has raised state-of-the-art performance almost by the figure of 8% in terms of F 1 measure, the RPC-snRC with character level features and one hot encoding manages to mark the peak performance at 95% thereby clearly outperforming all the other systems (previously existing systems and Resnet based systems adapted in this research).

In a nutshell, convolutional neural network based deep architectures have the ability to extract discriminative features directly from primary sequences of small non-coding RNA. his is depicted by the results where performances of the proposed and adapted methodologies are significantly higher than the state-of-the-art methodologies which take secondary structural features as input. Moreover, performance of ResNet based architectures is lower than the performance of the proposed RPC-snRC methodology because in ResNet models gradient does not flow properly from subsequent layers to previous layers [74]. It can also be inferred that ResNet model with 50 layers extracted some irrelevant and redundant features which slightly reduced its performance as coapared to the performance of ResNet 18 layers model.

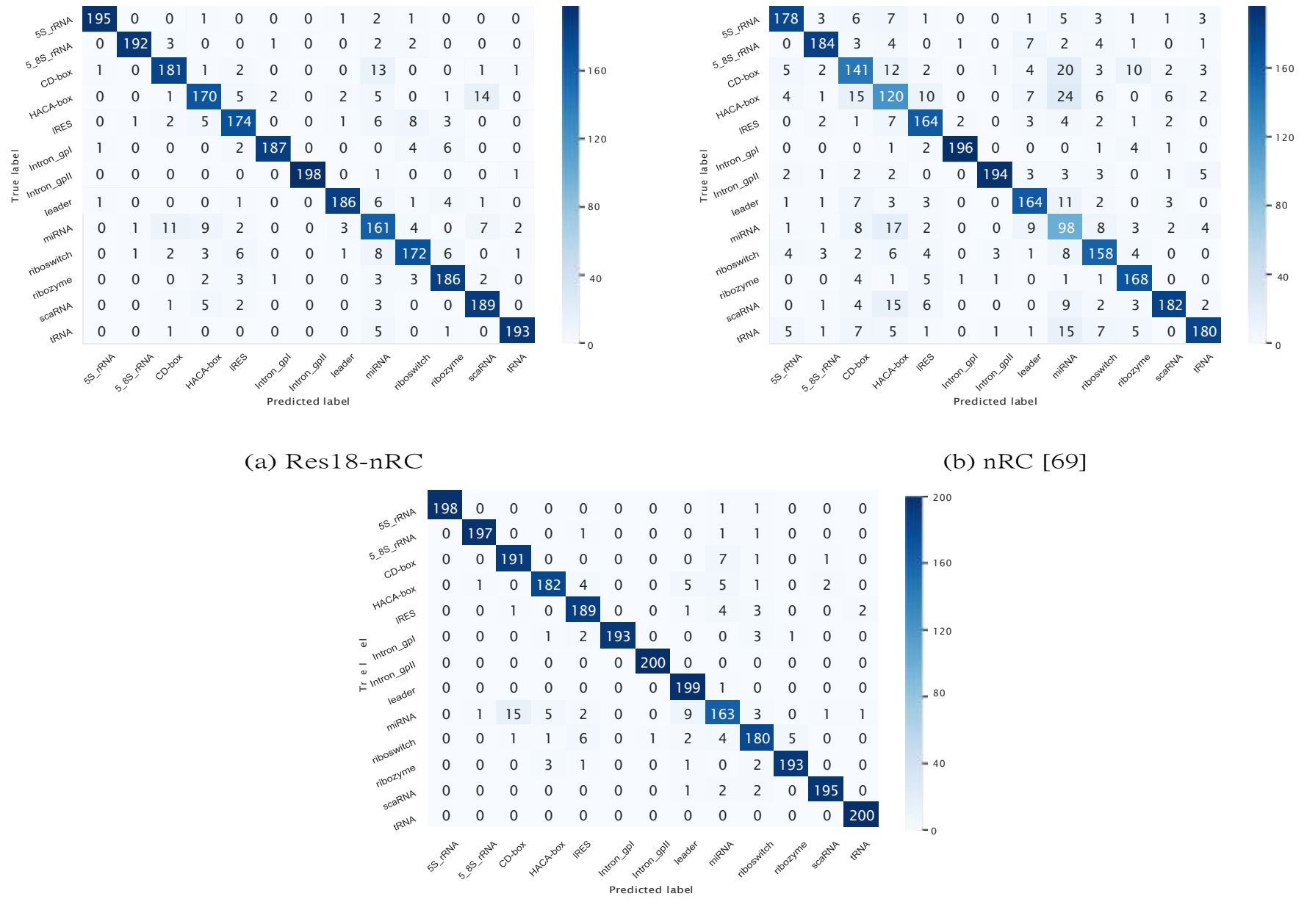

(a) Res18-nRC

(b) nRC [69]

(c) RPC-snRC

Figure 3: Accuracy Confusion matrix of the proposed RPC-snRC, Adapted Res18-nRC and state- of-the-art nRC [69] classiftcation methodologies.

In order to further compare the performance of the proposed RPC-snRC and the adapted ResNet based methodologies with the state-of-the-art methods, a class level performance comparison is performed in terms of accuracy confusion matrix. Accuracy confusion matrices of RPC-snRC, ResNet-18, and nRC methodologies on the test set of nCR dataset are shown in the Figure 3. Note that the RNAGCN [28] is the most recently reported method for small non-coding RNA classification, however, the authors have not provided class level results of their method. Therefore, we performed class level performance comparison of the proposed RPC-snRC and adapted methodologies with nRC classification methodology. The proposed RPC-snRC and the adapted Res18-snRC based methodologies produce highest performance with character level and one hot vector representation. So here we take confusion matrices of both methodologies with highest performance values. As depicted by Figure 3, RPC-snRC methodology correctly classifies all the 200 samples of two classes namely Intron gpII and tRNA as compared to the state-of-the-art nRC methodology which manages to correctly classify only 180 samples of tRNA and 196 samples of Intron gpII class. Performance of Res18-snRC remains in between the performance of nrc and RPC-snRC methodologies as it correctly predicted 198 samples of Intron gpII and 193 samples of tRNA clas. In addition, state-of-the-art nRC methodology fails to mark prominent performance as significant samples of almost every class are mistakenly classified in miRNA, HACA-box, CD-box, and IRES classes, while, only a few samples of each class are misclassified in the proposed RPC-snRC methodology. Although, miRNA has shown the lowest performance amongst all classes in both methodologies, however, the proposed RPC-snRC still correctly classifies 163 samples out of the maximum possible 200 as compared to state-of-the-art nRC methodology which only manages to correctly classify 98 samples. Also, the proposed RPC-snRC methodology successfully classifies more than 190 samples in each of the nine classes, i.e., introl gpll, tRNA, 5S rRNA, 5 8S rRNA, leader, scaRNA, ribozyme, introl gpl, and CD-box. Whereas, rest of the classes achieve the substantial count of 180's and 160's as shown by Figure3. Contrarily, in state-of-the-art nRC methodology, only two classes intron gpl, and intron gpll correctly classify more than 190 samples. Similarly, the adapted Res18-nRc methodology was able to correctly predict more than 190 samples for 4 classes, namely 5S rRNA, 5 8S rRNA, introl gpll, and tRNA.

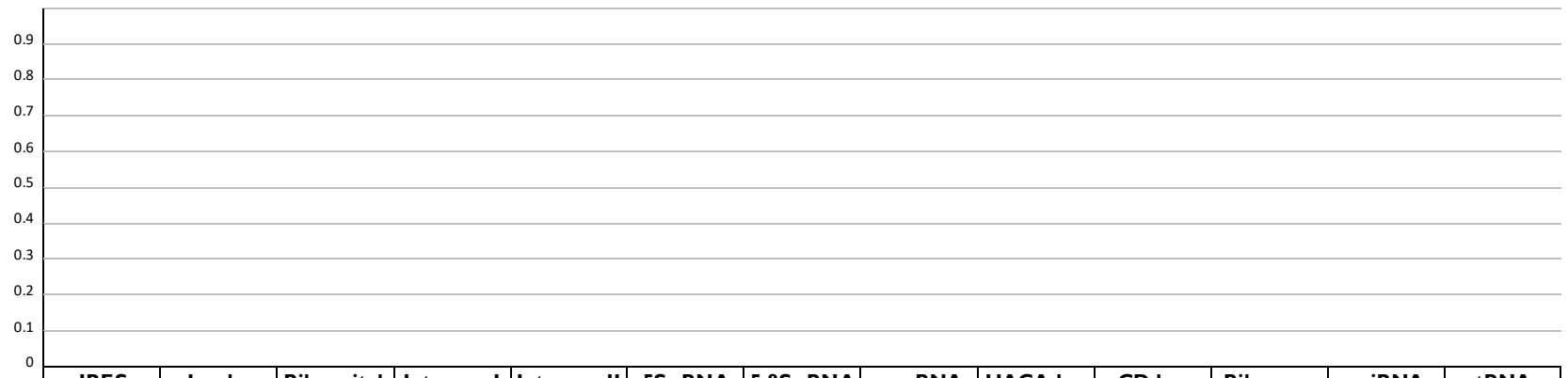

| | | IRES | Leader | Riboswitch | Intron gpI | Intron gpII | 5S rRNA | 5.8S rRNA | scaRNA | HACA-box | CD-box | Ribozyme | miRNA | tRNA |
|---|---|---|---|---|---|---|---|---|---|---|---|---|---|---|
| nRC | Precision | 0.860 | 0.844 | 0.643 | 0.563 | 0.870 | 0.954 | 0.868 | 0.809 | 0.585 | 0.739 | 0.889 | 0.748 | 0.746 |
| | Recall | 0.860 | 0.895 | 0.615 | 0.485 | 0.840 | 0.945 | 0.960 | 0.805 | 0.515 | 0.695 | 0.845 | 0.880 | 0.840 |
| | F1-Score | 0.860 | 0.868 | 0.6291 | 0.521 | 0.854 | 0.949 | 0.912 | 0.807 | 0.547 | 0.716 | 0.866 | 0.809 | 0.790 |
| Deep | Precision | 0.922 | 0.913 | 0.914 | 1.000 | 0.995 | 1.000 | 0.990 | 0.980 | 0.948 | 0.918 | 0.969 | 0.867 | 0.985 |
| | Recall | 0.945 | 0.995 | 0.900 | 0.965 | 1.000 | 0.990 | 0.985 | 0.975 | 0.910 | 0.955 | 0.965 | 0.815 | 1.000 |
| | F1-Score | 0.933 | 0.952 | 0.906 | 0.982 | 0.997 | 0.995 | 0.987 | 0.977 | 0.928 | 0.936 | 0.967 | 0.840 | 0.992 |

Figure 4: Individual class performance of proposed RPC-snRC methodology and state-of-the-art nRC [69] methodology on small ncRNA classiftcation dataset

Figure 4 shows individual class level performances of RPC-snRC and nRC classification methodologies over small ncRNA classification dataset in terms of precision, recall, and $F_1$ measure. Overall, for all classes, RPC-snRC methodology significantly outperforms the state-of-the-art nRC methodology in all three performance metrics except miRNA class, where nRC methodology manages to mark better recall figure. Moreover, amongst all performance metrices, nRC classification methodology manages to sustain performance values of precision, recall and $F_1$ measure only for three classes(IRES, 5 8S rRNA, scaRNA), on the other hand, the performance of RPC-snRC classification methodology stays consistent for 7 classes namely ribozymes, 5S rRNA, 5 8S rRNA, tRNA, scaRNA, Intron gpII, and riboswitch. This unique behaviour of RPC-snRC methodology shows that it suffers less from type I, and type II errors as compared to nRC methodology-performance of whom seems less stable at class level.

## 5. Conclusion

This paper proposes a novel methodology, named RPC-snRC, which classifies small non-coding RNA sequences into their relevant families by utilizing positioning and occurrences information of various nucleotides. Experimental results prove that the proposed RPC-snRC methodology is highly robust as it is neither biased towards false positive nor false negative predictions. Adapted Res18-snRC and Res50-snRC methodologies perform better than the state-of the-art small non-coding RNA classification methodologies. However, their performance is not as promising as produced by the proposed RPC-snRC methodology because in ResNet architectures gradient cannot flow properly from subsequent layers to previous layers. The proposed RPC-snRC methodology marks highest F1-score of 95% by representing character based features through one hot encoding as compared to state-of-the-art ncRNA, RNAGCn, and the adapted Res18-nRC, Res50-nRC classification methodologies which only manage to produce the performance of 77%, 85%, 91%, and 89% respectively. Moreover, in our experimentation, almost all methodologies perform better with one hot vector encoding than randomly initialized or pretrained word embeddings. From these results, it can be concluded that character or atom level feature produces better performance as compared to k-mers based features. A compelling future line of our work would be exploring the impact of a hybrid methodology which shall reap the benefits of both primary and secondary structural features.

## Supplementary material

To reproduce the results Source code and all other data will be publicly available.

## References

[1] W. Kraczkowska, P. P. Jagodziński, The long non-coding rna landscape of atherosclerotic plaques, Molecular diagnosis & therapy (2019) 1–15.

[2] H. Ghasemi, Z. Sabati, H. Ghaedi, Z. Salehi, B. Alipoor, Circular rnas in β-cell function and type 2 diabetes-related complications: a potential diagnostic and therapeutic approach, Molecular biology reports (2019) 1–13.

[3] F. Hubé, C. Francastel, Coding and non-coding rnas, the frontier has never been so blurred, Frontiers in Genetics 9 (2018) 140.

[4] M. E. Dinger, K. C. Pang, T. R. Mercer, J. S. Mattick, Differentiating protein-coding and noncoding rna: challenges and ambiguities, PLoS computational biology 4 (11) (2008) e1000176.

[5] N. Amin, A. McGrath, Y.-P. P. Chen, Evaluation of deep learning in non-coding rna classification, Nature Machine Intelligence 1 (5) (2019) 246.

[6] Y. Fang, M. J. Fullwood, Roles, functions, and mechanisms of long non-coding rnas in cancer, Genomics, proteomics & bioinformatics 14 (1) (2016) 42 – 54.

[7] N. Yu, Z. Yu, Y. Pan, A deep learning method for lincrna detection using auto-encoder algorithm, BMC bioinformatics 18 (15) (2017) 511.

[8] M. Chaabane, End-to-end learning framework for circular rna classification from other long non-coding rnas using multi-modal deep learning.

[9] Y. Ma, X. Zhang, Y.-Z. Wang, H. Tian, S. Xu, Research progress of circular rnas in lung cancer, Cancer biology & therapy 20 (2) (2019) 123 – 129.

[10] T. Li, S. Wang, R. Wu, X. Zhou, D. Zhu, Y. Zhang, Identification of long non-protein coding rnas in chicken skeletal muscle using next generation sequencing, Genomics 99 (5) (2012) 292 – 298.

[11] Y.-J. Kang, D.-C. Yang, L. Kong, M. Hou, Y.-Q. Meng, L. Wei, G. Gao, Cpc2: a fast and accurate coding potential calculator based on sequence intrinsic features, Nucleic acids research 45 (W1) (2017) W12 – W16.

[12] J. Baek, B. Lee, S. Kwon, S. Yoon, Lncrnanet: long non-coding rna identification using deep learning, Bioinformatics 34 (22) (2018) 3889 – 3897.

[13] C. Yang, L. Yang, M. Zhou, H. Xie, C. Zhang, M. D. Wang, H. Zhu, Lncadeep: an ab initio lncrna identification and functional annotation tool based on deep learning, Bioinformatics 34 (22) (2018) 3825 – 3834.

[14] S. Han, Y. Liang, Q. Ma, Y. Xu, Y. Zhang, W. Du, C. Wang, Y. Li, Lncfinder: an integrated platform for long non-coding rna identification

utilizing sequence intrinsic composition, structural information and physicochemical property, Briefings in bioinformatics.

[15] G. Stefani, F. J. Slack, Small non-coding rnas in animal development, Nature reviews Molecular cell biology 9 (3) (2008) 219.

[16] K. Wang, D. Singh, Z. Zeng, S. J. Coleman, Y. Huang, G. L. Savich, X. He, P. Mieczkowski, S. A. Grimm, C. M. Perou, et al., Mapsplice: accurate mapping of rna-seq reads for splice junction discovery, Nucleic acids research 38 (18) (2010) e178 – e178.

[17] M. Jovanovic, M. Hengartner, mirnas and apoptosis: Rnas to die for, Oncogene 25 (46) (2006) 6176.

[18] I. Büssing, F. J. Slack, H. Großhans, let-7 micrornas in development, stem cells and cancer, Trends in molecular medicine 14 (9) (2008) 400 – 409.

[19] R. Schickel, B. Boyerinas, S. Park, M. Peter, Micrornas: key players in the immune system, differentiation, tumorigenesis and cell death, Oncogene 27 (45) (2008) 5959.

[20] B. Hrdlickova, R. C. de Almeida, Z. Borek, S. Withoff, Genetic variation in the non-coding genome: Involvement of micro-rnas and long non-coding rnas in disease, Biochimica et Biophysica Acta (BBA)-Molecular Basis of Disease 1842 (10) (2014) 1910 – 1922.

[21] M. Esteller, Non-coding rnas in human disease, Nature reviews genetics 12 (12) (2011) 861.

[22] D. Ammons, J. Rampersad, G. E. Fox, 5s rrna gene deletions cause an unexpectedly high fitness loss in escherichia coli, Nucleic acids research 27 (2) (1999) 637 – 642.

[23] S. A. Elela, R. N. Nazar, Role of the 5.8 s rrna in ribosome translocation, Nucleic acids research 25 (9) (1997) 1788 – 1794.

[24] B. Fontoura, C. A. Atienza, E. A. Sorokina, T. Morimoto, R. B. Carroll, Cytoplasmic p53 polypeptide is associated with ribosomes., Molecular and cellular biology 17 (6) (1997) 3146 – 3154.

[25] R. Shi, V. L. Chiang, Facile means for quantifying microrna expression by real-time pcr, Biotechniques 39 (4) (2005) 519 – 525.

[26] R. N. Nazar, The ribosomal 5.8 s rna: eukaryotic adaptation or processing variant?, Canadian journal of biochemistry and cell biology 62 (6) (1984) 311 – 320.

[27] A. Fiannaca, M. La Rosa, L. La Paglia, R. Rizzo, A. Urso, nrc: non-coding rna classifier based on structural features, BioData mining 10 (1) (2017) 27.

[28] E. Rossi, F. Monti, M. Bronstein, P. Liò, ncrna classification with graph convolutional networks, arXiv preprint arXiv:1905.06515.

[29] X. Fu, W. Zhu, L. Cai, B. Liao, L. Peng, Y. Chen, J. Yang, Improved pre-mirnas identification through mutual information of pre-mirna sequences and structures, Frontiers in genetics 10.

[30] S. Washietl, I. L. Hofacker, P. F. Stadler, Fast and reliable prediction of noncoding rnas, Proceedings of the National Academy of Sciences 102 (7) (2005) 2454 – 2459.

[31] S. Griffiths-Jones, A. Bateman, M. Marshall, A. Khanna, S. R. Eddy, Rfam: an rna family database, Nucleic acids research 31 (1) (2003) 439 – 441.

[32] S. Washietl, I. L. Hofacker, Consensus folding of aligned sequences as a new measure for the detection of functional rnas by comparative genomics, Journal of molecular biology 342 (1) (2004) 19 – 30.

[33] I. L. Hofacker, M. Fekete, P. F. Stadler, Secondary structure prediction for aligned rna sequences, Journal of molecular biology 319 (5) (2002) 1059 – 1066.

[34] J. Liu, J. Gough, B. Rost, Distinguishing protein-coding from non-coding rnas through support vector machines, PLoS genetics 2 (4) (2006) e29.

[35] P. Carninci, T. Kasukawa, S. Katayama, J. Gough, M. Frith, N. Maeda, R. Oyama, T. Ravasi, B. Lenhard, C. Wells, et al., The transcriptional landscape of the mammalian genome, Science 309 (5740) (2005) 1559 – 1563.

[36] C. Liu, B. Bai, G. Skogerbø, L. Cai, W. Deng, Y. Zhang, D. Bu, Y. Zhao, R. Chen, Noncode: an integrated knowledge database of non-coding rnas, Nucleic acids research 33 (suppl_1) (2005) D112 – D115.

[37] K. C. Pang, S. Stephen, P. G. Engström, K. Tajul-Arifin, W. Chen, C. Wahlestedt, B. Lenhard, Y. Hayashizaki, J. S. Mattick, Rnadb—a comprehensive mammalian noncoding rna database, Nucleic acids research 33 (suppl_1) (2005) D125 – D130.

[38] S. Lertampaiporn, C. Thammarongtham, C. Nukoolkit, B. Kaewkamnerdpong, M. Ruengjitchatchawalya, Identification of non-coding rnas with a new composite feature in the hybrid random forest ensemble algorithm, Nucleic acids research 42 (11) (2014) e93 – e93.

[39] K. D. Pruitt, T. Tatusova, G. R. Brown, D. R. Maglott, Ncbi reference sequences (refseq): current status, new features and genome annotation policy, Nucleic acids research 40 (D1) (2011) D130 – D135.

[40] P. P. Amaral, M. B. Clark, D. K. Gascoigne, M. E. Dinger, J. S. Mattick, lncrnadb: a reference database for long noncoding rnas, Nucleic acids research 39 (suppl_1) (2010) D146 – D151.

[41] S. F. Altschul, W. Gish, W. Miller, E. W. Myers, D. J. Lipman, Basic local alignment search tool, Journal of molecular biology 215 (3) (1990) 403 – 410.

[42] R. Achawanantakun, J. Chen, Y. Sun, Y. Zhang, Lncrna-id: Long non-coding rna identification using balanced random forests, Bioinformatics 31 (24) (2015) 3897 – 3905.

[43] G. Chen, Z. Wang, D. Wang, C. Qiu, M. Liu, X. Chen, Q. Zhang, G. Yan, Q. Cui, Lncrnadisease: a database for long-non-coding rna-associated diseases, Nucleic acids research 41 (D1) (2012) D983 – D986.

[44] Y. Saito, K. Sato, Y. Sakakibara, Fast and accurate clustering of noncoding rnas using ensembles of sequence alignments and secondary structures, BMC bioinformatics 12 (1) (2011) S48.

[45] J. H. Havgaard, E. Torarinsson, J. Gorodkin, Fast pairwise structural rna alignments by pruning of the dynamical programming matrix, PLOS computational biology 3 (10) (2007) e193.

[46] K. Sato, T. Mituyama, K. Asai, Y. Sakakibara, Directed acyclic graph kernels for structural rna analysis, BMC bioinformatics 9 (1) (2008) 318.

[47] S. Will, K. Reiche, I. L. Hofacker, P. F. Stadler, R. Backofen, Inferring noncoding rna families and classes by means of genome-scale structure-based clustering, PLoS computational biology 3 (4) (2007) e65.

[48] M. Miladi, A. Junge, F. Costa, S. E. Seemann, J. H. Havgaard, J. Gorodkin, R. Backofen, Rnascclust: clustering rna sequences using structure conservation and graph based motifs, Bioinformatics 33 (14) (2017) 2089 – 2096.

[49] M. Miladi, A. Junge, F. Costa, S. E. Seemann, J. H. Havgaard, J. Gorodkin, R. Backofen, Rnascclust: clustering rna sequences using structure conservation and graph based motifs, Bioinformatics 33 (14) (2017) 2089 – 2096.

[50] G. Aoki, Y. Sakakibara, Convolutional neural networks for classification of alignments of non-coding rna sequences, Bioinformatics 34 (13) (2018) i237 – i244.

[51] A. Fiannaca, M. La Rosa, L. La Paglia, R. Rizzo, A. Urso, nrc: non-coding rna classifier based on structural features, BioData mining 10 (1) (2017) 27.

[52] J. Baek, B. Lee, S. Kwon, S. Yoon, Lncrnanet: long non-coding rna identification using deep learning, Bioinformatics 34 (22) (2018) 3889 – 3897.

[53] S. Park, S. Min, H.-S. Choi, S. Yoon, Deep recurrent neural network-based identification of precursor micrornas, in: Advances in Neural Information Processing Systems, 2017, pp. 2891 – 2900.

[54] M. Tsuchiya, K. Amano, M. Abe, M. Seki, S. Hase, K. Sato, Y. Sakakibara, Sharaku: an algorithm for aligning and clustering read mapping profiles of deep sequencing in non-coding rna processing, Bioinformatics 32 (12) (2016) i369 – i377.

[55] K. Sato, Y. Kato, T. Akutsu, K. Asai, Y. Sakakibara, Dafs: simultaneous aligning and folding of rna sequences via dual decomposition, Bioinformatics 28 (24) (2012) 3218 – 3224.

[56] J. S. McCaskill, The equilibrium partition function and base pair binding probabilities for rna secondary structure, Biopolymers: Original Research on Biomolecules 29 (6-7) (1990) 1105 – 1119.

[57] R. Weikard, F. Hadlich, C. Kuehn, Identification of novel transcripts and noncoding rnas in bovine skin by deep next generation sequencing, BMC genomics 14 (1) (2013) 789.

[58] S. Washietl, I. L. Hofacker, P. F. Stadler, Fast and reliable prediction of noncoding rnas, Proceedings of the National Academy of Sciences 102 (7) (2005) 2454 – 2459.

[59] J. Liu, J. Gough, B. Rost, Distinguishing protein-coding from non-coding rnas through support vector machines, PLoS genetics 2 (4) (2006) e29.

[60] S. Lertampaiporn, C. Thammarongtham, C. Nukoolkit, B. Kaewkamnerdpong, M. Ruengjitchatchawalya, Identification of non-coding rnas with a new composite feature in the hybrid random forest ensemble algorithm, Nucleic acids research 42 (11) (2014) e93 – e93.

[61] S. Park, S. Min, H.-S. Choi, S. Yoon, Deep recurrent neural network-based identification of precursor micrornas, in: Advances in Neural Information Processing Systems, 2017, pp. 2891 – 2900.

[62] J. Baek, B. Lee, S. Kwon, S. Yoon, Lncrnanet: long non-coding rna identification using deep learning, Bioinformatics 34 (22) (2018) 3889 – 3897.

[63] Y. Saito, K. Sato, Y. Sakakibara, Fast and accurate clustering of noncoding rnas using ensembles of sequence alignments and secondary structures, BMC bioinformatics 12 (1) (2011) S48.

[64] M. Miladi, A. Junge, F. Costa, S. E. Seemann, J. H. Havgaard, J. Gorodkin, R. Backofen, Rnascclust: clustering rna sequences using structure conservation and graph based motifs, Bioinformatics 33 (14) (2017) 2089 – 2096.

[65] M. Tsuchiya, K. Amano, M. Abe, M. Seki, S. Hase, K. Sato, Y. Sakakibara, Sharaku: an algorithm for aligning and clustering read mapping profiles of deep sequencing in non-coding rna processing, Bioinformatics 32 (12) (2016) i369 – i377.

[66] G. Aoki, Y. Sakakibara, Convolutional neural networks for classification of alignments of non-coding rna sequences, Bioinformatics 34 (13) (2018) i237 – i244.

[67] R. Weikard, F. Hadlich, C. Kuehn, Identification of novel transcripts and noncoding rnas in bovine skin by deep next generation sequencing, BMC genomics 14 (1) (2013) 789.

[68] M. Chaabane, End-to-end learning framework for circular rna classification from other long non-coding rnas using multi-modal deep learning.

[69] S. Memczak, M. Jens, A. Elefsinioti, F. Torti, J. Krueger, A. Rybak, L. Maier, S. D. Mackowiak, L. H. Gregersen, M. Munschauer, et al., Circular rnas are a large class of animal rnas with regulatory potency, Nature 495 (7441) (2013) 333.

[70] A. Krizhevsky, I. Sutskever, G. E. Hinton, Imagenet classification with deep convolutional neural networks, in: Advances in neural information processing systems, 2012, pp. 1097 – 1105.

[71] K. Simonyan, A. Zisserman, Very deep convolutional networks for large-scale image recognition, arXiv preprint arXiv:1409.1556.

[72] C. Szegedy, W. Liu, Y. Jia, P. Sermanet, S. Reed, D. Anguelov, D. Erhan, V. Vanhoucke, A. Rabinovich, Going deeper with convolutions, in: Proceedings of the IEEE conference on computer vision and pattern recognition, 2015, pp. 1 – 9.

[73] K. He, X. Zhang, S. Ren, J. Sun, Deep residual learning for image recognition, in: Proceedings of the IEEE conference on computer vision and pattern recognition, 2016, pp. 770 – 778.

[74] G. Huang, Z. Liu, L. Van Der Maaten, K. Q. Weinberger, Densely connected convolutional networks, in: Proceedings of the IEEE conference on computer vision and pattern recognition, 2017, pp. 4700 – 4708.

[75] A. Fabijańska, S. Grabowski, Viral genome deep classifier, IEEE Access 7 (2019) 81297 – 81307.

[76] E. Asgari, M. R. Mofrad, Continuous distributed representation of biological sequences for deep proteomics and genomics, PloS one 10 (11) (2015) e0141287.

[77] S. Ioffe, C. Szegedy, Batch normalization: Accelerating deep network training by reducing internal covariate shift, arXiv preprint arXiv:1502.03167.

[78] X. Glorot, A. Bordes, Y. Bengio, Deep sparse rectifier neural networks, in: Proceedings of the fourteenth international conference on artificial intelligence and statistics, 2011, pp. 315 – 323.

[79] Y. LeCun, L. Bottou, Y. Bengio, P. Haffner, et al., Gradient-based learning applied to document recognition, Proceedings of the IEEE 86 (11) (1998) 2278 – 2324.

[80] V. Subramanian, Deep Learning with PyTorch: A practical approach to building neural network models using PyTorch, Packt Publishing Ltd, 2018.

[81] D. P. Kingma, J. Ba, Adam: A method for stochastic optimization, arXiv preprint arXiv:1412.6980.